\documentclass[prc,showpacs,floatfix]{revtex4}
\usepackage{bm}
\usepackage{dcolumn}
\usepackage{graphicx}
\begin{document}
\preprint{MKPH-T-05-10}
\title{
{\bf Binding of charmonium with two- and three-body nuclei}}
\author{V.\,B. Belyaev,
N.\,V. Shevchenko\footnote{Present address: Nuclear Physics Institute, 25068 \v{R}e\v{z}, Czech Republic}
}
\affiliation{Joint Institute for Nuclear Research, Dubna, 141980,
Russia}
\author{A. Fix}
\affiliation{ Institut f\"ur Kernphysik, Johannes
Gutenberg-Universit\"at Mainz, D-55099 Mainz, Germany}
\author{W. Sandhas}
\affiliation{Physikalisches Institut, Universit\H{a}t Bonn, D-53115
Bonn, Germany}
\date{\today}
\begin{abstract}
The energies of the $\eta_c d$ and $\eta_c\,^3$He bound states are
calculated on the basis of exact three- and four-body AGS equations.
For the $\eta_c N$ interaction a Yukawa-type potential has been
adopted. The calculations are done for a certain range of its
strength parameter. The results obtained are quite different from
calculations based on the folding model.
\end{abstract}

 \maketitle

\section{Introduction}\label{intr}

The behavior of mesons in a nuclear medium has attracted much
attention during the last years \cite{Mosel}. We mention the
existence and properties of deeply bound meson-nuclear states
\cite{Toki} as well as the possible restoration of chiral symmetry
in the nuclear environment \cite{Tsuchima}, a point closely related
to modifications of widths and masses of the mesons propagating
through nuclear matter \cite{Keiser}.

It should be emphasized that all these phenomena are usually
described in terms of effective degrees of freedom including
mesons, nucleons, and isobars. The manifestation of more
fundamental quark-gluon degrees of freedom at low energies is
very unlikely because of the strong mismatch between the nuclear
energy scale (few tenth of MeV) and the QCD energy scale (few
hundreds or thousands of MeV).

However, as pointed out in \cite{Brodsky} with regard to
charmonium-like mesons, for example $\eta_c$, which do not contain
$u$- and $d$-quarks, the main contribution to the meson-nucleon
interaction is expected to originate not from effective, but from
fundamental QCD degrees of freedom, namely from the few-gluon
exchange. In this way, using the Pomeron exchange model, one can
deduce a Yukawa-type potential for the $\eta_c$-nucleon system
\cite{Brodsky}. Following these ideas, the authors of \cite{Teram}
were able to explain the $pp$-spin correlation at the threshold
energy of charm production. Such an analysis allows one to restrict
the uncertainties of the parameters of the charmonium-nucleon
potential. Independent theoretical indication on the existence of
attraction between charmonium and nucleons has been obtained in
\cite{Luke}. In case of an infinite-mass limit of the heavy quark,
these authors have obtained a binding energy of J/$\psi$ in nuclear
matter of the order of 10 MeV. For the (2S) charmonium state this
approach predicts an even stronger binding. Despite the above
indications, the data of the charmonium-$N$ interaction are still
rather scarce. Bearing this in mind, it appears reasonable to look
for possible formations of the corresponding mesic nuclei, similar
to the $\eta$-nuclear systems which were extensively discussed in
the literature (see e.g.~\cite{ours} and references therein). In
particular the study of the reactions
\begin{equation}\label{20}
\bar{p}+A \to \bar{p}+p+A' \to \eta_c+A'
\end{equation}
has been planned within the PANDA project \cite{panda}. The
production of $\eta_c$ in electromagnetic processes has already been
started in the CLIO collaboration \cite{clio}. As explained in
\cite{Brodsky}, the precise knowledge of the binding energy of
meson-nuclear systems is of crucial importance for their
experimental observation.

First attempts to calculate the binding of $\eta_c$-mesons with
light nuclei have been made in Ref.~\cite{Brodsky} on the basis of
variational calculations, and in Ref.~\cite{Wasson} where the
folding model was used. Both calculations involve a Yukawa form for
the basic $\eta_c N$ interaction. Our results for the binding
energies of $\eta_c d$ and $\eta_c\,^3$He, presented below, are
obtained within the AGS theory \cite{AGS,GS}. Similar to
\cite{Brodsky,Wasson} we use the Yukawa type $\eta_{c} N$ potential
\begin{equation}\label{60}
V_{\eta_c N}(r) = - a\,\frac{e^{-\alpha r}}{r}\,
\end{equation}
with $\alpha$ = 0.6 GeV. The parameter $a$ is varied between 0.4 and
0.6. Since the two-body potentials act essentially via their
$s$-waves contributions, we take into account only $s$-wave parts of
the corresponding $\eta_cN$ and $NN$ scattering amplitudes entering
the kernel of the AGS equations.

\section{Formalism}\label{formal}

Let us start by giving a brief description of the 3-body
$\eta_{c}NN$ formalism. Its main ingredient is a separable
representation of the driving $\eta_c N$ and $NN$ potentials
\begin{equation}\label{30}
V_\alpha=\sum\limits_{i,j}^N|\chi^i_\alpha\rangle\langle\chi^j_\alpha|\,,
\quad \alpha\in\{\eta_cN,\,NN\}\,.
\end{equation}
To outline the important aspects of this approach, we use for the
moment an oversimplified representation of the $\eta_c N$
potential keeping only the first term in the sum (\ref{30}), while
the actual calculations are performed with a more realistic
representation containing six terms (we will discuss the
sensitivity of our results to the number of terms kept in the
separable expansion). For the $s$-wave triplet state $^3S_1$ of
two nucleons we employ the separable rank-one version of the
Paris potential of Ref.~\cite{Zankel}.

The AGS equations for the $\eta_c NN$ system read \cite{ours}
\begin{eqnarray}\label{40}
X_{11}(z) &=& 2 \, Z_{21}(z) \, \tau_2(z-\frac{p_2^2}{2 \mu_2}) \,
X_{21}(z) \\
\nonumber X_{21}(z) &=& Z_{21} + Z_{21}(z) \,
\tau_1(z-\frac{p_1^2}{2 \mu_1}) \, X_{11}(z) + Z_{23}(z) \,
\tau_2(z-\frac{p_2^2}{2 \mu_2}) \, X_{21}(z)\,.
\end{eqnarray}
The functions $\tau_\alpha(z)$, $Z_{\alpha\beta}$, and
$X_{\alpha\beta}$ ($\alpha,\beta$=1,2) can be expressed in terms of
the AGS transition operators $U_{\beta\alpha}$ and form factors
$|\chi_{\alpha}\rangle$ appearing in (\ref{30}) via
\begin{eqnarray}\label{50}
Z_{\beta \alpha}(z) &\equiv& (1-\delta_{\alpha \beta}) \langle
\chi_{\beta} |
G_0(z) | \chi_{\alpha} \rangle \\
X_{\beta \alpha}(z) &\equiv& \langle \chi_{\beta} | G_0(z) \,
U_{\beta \alpha} \, G_0(z) | \chi_{\alpha} \rangle
\end{eqnarray}
where $G_0(z)$ is the free $\eta_c NN$ propagator. As mentioned, for
the $NN$ subsystem we use a separable representation of the Paris
potential
\begin{equation}\label{140}
V_{NN}(k,k')=-g(k)g(k')\,,\quad \mbox{where}\
g(k)=\sum_{i=1}^6\frac{C_i}{k^2+\beta_i^2}\,,
\end{equation}
with parameters $C_i$ and $\beta_i$ listed for the $^3S_1$
configuration in \cite{Zankel}.

For the $\eta_cN$ interaction we first consider the rank-one
representation of the $\eta_cN$ potential (\ref{30}) within the
Bateman method \cite{VB}. The corresponding general expression reads
\begin{equation}\label{70}
V_0^{[N]}(k,k') = \sum_{i,j}^{N} V_0(k,s_i) \, d^{-1}_{ij} \,
V_0(s_j,k'),
\end{equation}
with $d_{ij} = V_0(s_i,s_j)$. Here $s_i$ is an expansion parameter
and
\begin{equation}\label{80}
V_0(k,k') = \frac{1}{2 \pi^2} \, \int_{0}^{\infty} j_0(kr) \, V(r)
\, j_0(k' r) \, r^2 \, dr
\end{equation}
the $s$-wave Fourier transform of the potential $V(r)$. Using
expression~(\ref{60}) for the potential $V(r)$ in the
integral~(\ref{80}) and taking $N=1$ and $s_1=0$ in the
expansion~(\ref{70}) one can readily obtain
\begin{equation}\label{90}
V_0^{[1]}(k,k') = \frac{V(k,0) V(0,k')}{V(0,0)} = -4 \pi a
\frac{\alpha^2}{(k^2+\alpha^2)(k'^2+\alpha^2)}\,.
\end{equation}

If the Bateman method is applied to a negative potential, the
next terms in the separable expansion tend to increase the
attraction. Therefore, the neglect of the higher order terms in
the sum~(\ref{70}) will result in underestimation of the absolute
value of the $\eta_c d$ binding energy. An obvious advantage of
the Bateman method is that it provides analytical expressions for
all terms in (\ref{70}).

It is worthwhile to note that the potential (\ref{90}) predicts an
antibound (virtual) state in the $\eta_c N$ system lying at
\begin{equation}\label{100}
\epsilon=-\frac{1}{2\mu}\left(\sqrt{\frac{a\mu\alpha^3}
{4\pi}}-\alpha\right)^2=-6.1\ \mbox{MeV}\,,
\end{equation}
where $\mu$ is $\eta_cN$ reduced mass. The corresponding
scattering length
\begin{equation}\label{110}
a_{\eta_c N}=\frac{2}{\frac{4\pi}{a\mu}-\alpha}\approx 0.97\
\mbox{fm}
\end{equation}
is a bit larger than the $\eta N$ scattering length predicted by the
modern analyses (see e.g.\ compilation in \cite{Wycech}).
Furthermore, there is no absorptive part in the $\eta_c N$
potential, which in the $\eta N$ case effectively increases the
repulsive properties of the interaction. The most important
difference between $\eta N$ and $\eta_c N$ is the larger mass of the
$\eta_c$ meson, which should increase the role of attraction in the
$\eta_c$ few-nucleon dynamics. According to this, one can expect
that the $\eta_c$ meson should be more tightly bound to the nucleons
than the $\eta$ meson. The energy of the $\eta_c d$ bound state,
calculated with the potential (\ref{90}) as a function of the
strength parameter $a$, is presented in Fig.~1 by the dashed line.

\begin{figure}
\includegraphics[width=8cm]{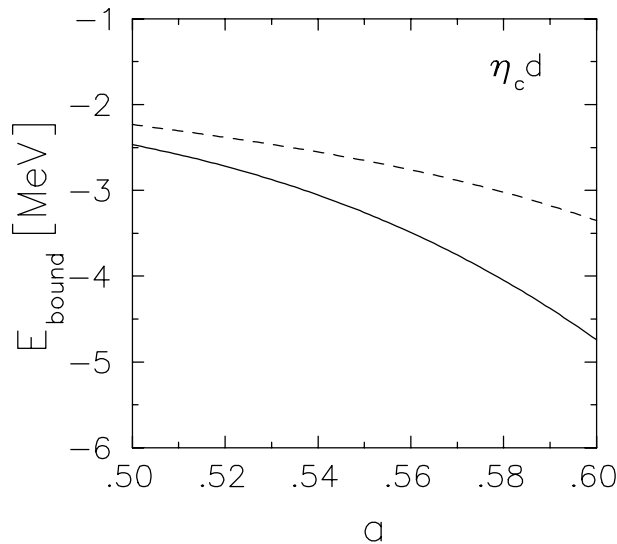}
\caption{Binding energy of $\eta_c d$ system, as function of the
strength $a$ of the Yukawa potential (\protect\ref{60}). The results
obtained with the Bateman and Hilbert-Schmidt representation of the
$\eta_cN$ potential are presented by the dashed and the solid line,
respectively.}
\end{figure}

As a next step we employ the Hilbert-Schmidt method for the
separable representation of the potential (\ref{60}). The
corresponding expansion reads
\begin{equation}\label{120}
V_{\eta_cN}(k,k')=\sum\limits_{n=1}^N\lambda_n v_n(k)v_n(k')\,,
\end{equation}
where the form factors $v_n(k)$ and strength parameters
$\lambda_n$ obey the homogeneous integral equation
\begin{equation}\label{130}
v_n(k)=\frac{1}{\lambda_n}\int\limits_0^\infty V_{\eta_c N}(k,k')
\frac{v_n(k')}{B-\frac{{k'}^2}{2\mu}}\frac{{k'}^2dk'}{2\pi^2}\,.
\end{equation}
$B$ = 0 was chosen in the actual calculation.

\section{Results}\label{result}

The $\eta_c d$ binding energy calculated for the
potential~(\ref{120}) with $N$ = 6 terms is presented in Fig.~1. As
expected, the last model predicts stronger binding than the leading
term in the Bateman expansion (\ref{70}). Furthermore as one can
see, in contrast to the results of \cite{Brodsky} and \cite{Wasson},
we find only one bound state lying in the region $-2$ MeV $\le E_b
\le -5$ MeV, when the parameter $a$ is varied between 0.5 and 0.6
GeV.

\begin{table}[!ht]
\renewcommand{\arraystretch}{2.0}
\caption{Dependence of the $\eta_c$d binding energy $E_b$ on the
number of separable terms $N$ in the expansion (\protect\ref{120}).
Parameter $a$ is equal to $0.55$.}
\begin{ruledtabular}
\begin{tabular}{c|cccc}
$N$ & 1 & 2 & 3 & 6 \\
\colrule
 $E_b$ [MeV] & -3.429 & -3.251 & -3.217   & -3.196 \\
\end{tabular}
\end{ruledtabular}
\end{table}

The results in Table~I show the accuracy of the Hilbert-Schmidt
method for the $\eta_c d$ binding energy. Evidently, already the
first two terms in the sum (\ref{120}) provide quite a satisfactory
approximation. Going up to $N$ = 6 results in only 1.7\,$\%$
increase of $|E_b|$.

Now we turn to the four-body $\eta_c\,^3$He system. For the
calculation we use the formalism developed in Ref.~\cite{Fix1} and
Ref.~\cite{Fix2} for the interaction of isoscalar-pseudoscalar
mesons with three-body nuclei. Since the formal part of the
problem is quite involved we do not present it here and refer the
reader to these references. It should only be recalled that the
calculations are based on separable representations of the two-
and three-body kernels of the basic AGS equations so that, as a
consequence, the resulting integral equations have the same
structure as the three-body equations (\ref{20}).

\begin{table}[!ht]
\renewcommand{\arraystretch}{2.0}
\caption{The binding energy of $\eta_c\,^3$He system (in MeV).}
\begin{ruledtabular}
\begin{tabular}{c|ccc}
$a$ & 4-Body calculations &
Variational calculation\protect\cite{Brodsky} &
Folding model \protect\cite{Wasson} \\
\colrule
0.4 & -1.3 & -3.0 & \\
0.6 & -14.5 & -19.0 & -0.8
\end{tabular}
\end{ruledtabular}
\end{table}

Let us compare our results for the binding energy of $\eta_c\,^3$He
system with variational \cite{Brodsky} and folding model
calculations \cite{Wasson}. One should emphasize that the
variational method used in \cite{Brodsky} is not an {\it ab initio}
many body treatment. Rather it is a model of meson-nuclear
interaction, postulating a Yukawa form for the meson-nuclear
potential with a range depending on the number of nucleons in the
nucleus. This model, as well as the folding model, takes into
account the important effect of smearing the meson-nucleon
interaction over the nuclear volume. However, the multiple
scattering of mesons in the nucleus is not accurately accounted for.
Our results of the microscopic four-body calculations of the
$\eta_c\,^3$He binding energy are presented in Fig.~2 and compared
with the alternative ones in Table~II. The calculations are
performed with the value $\alpha$ = 0.6 GeV for the range parameter
of the $\eta_cN$ potential.

\begin{figure}[!ht]
\includegraphics[width=8cm]{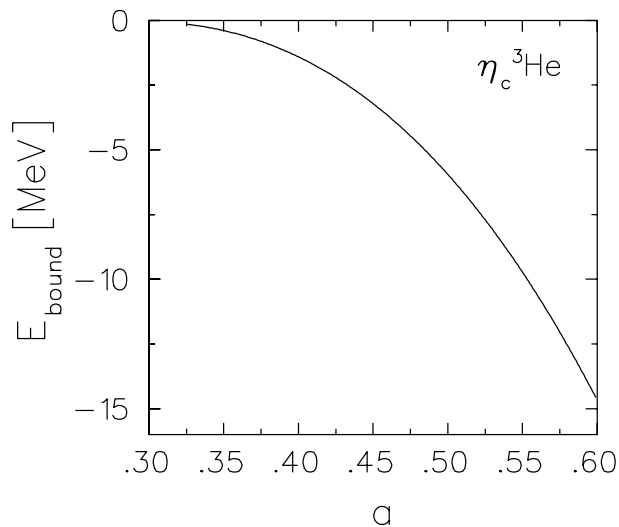}
\caption{Dependence of the $\eta_c\,^3$He binding energy on the
parameter $a$ in (\protect\ref{60}).}
\end{figure}

In summary, we calculated the binding energies of $\eta_c d$ and
$\eta_c\,^3$He using three- and four-body equations of the AGS
theory. The results exhibit essential differences to the predictions
of the folding model as well as to the variational calculations. The
characteristic values of the binding energies are about $-3$ MeV for
the $\eta_c d$ system and amount to $-14.5$ MeV for $a$ = 0.6 in the
$\eta_c\,^3$He case.

\section*{Acknowledgment}
This work was supported by the Deutsche Forschungsgemeinschaft and
the Russian Foundation for Basic Research, project 436RUS113/761.

\end{document}